\documentclass[11pt,showkeys]{revtex4-1}

\usepackage{amsmath}
\usepackage{amsfonts}
\usepackage{amssymb}
\usepackage[pdftex]{graphicx}
\usepackage{color}

\begin{document}
\title{Fermionization, Number of Families}
\author{N.S. Manko\v c Bor\v stnik}
\affiliation{Department of Physics, University of Ljubljana,
SI-1000 Ljubljana, Slovenia}
\author{H.B.F. Nielsen}
\thanks{H.B. Nielsen presented the talk.}
\affiliation{Niels Bohr Institute, University of Copenhagen, 
Blegdamsvej 17, Copenhagen {\O}, Denmark}

\begin{abstract}
 We investigate bosonization/fermionization for free massless
 fermions being equivalent to free massless bosons with the
 purpose of checking and correcting the old rule by Aratyn and
 one of us (H.B.F.N.)
 for the number of boson species relative to the number of
 fermion species which
 is required to have bosonization possible. An important
 application
 of such a counting of degrees of freedom relation would be to
 invoke restrictions
 on the number of families that could be possible under the
 assumption, that
 all the fermions in nature are the result of fermionizing a
 system of
 boson species. Since a theory of fundamental fermions can be
 accused for not being properly local because of having
 anticommutativity
 at space like distances rather than commutation as is more
 physically
 reasonable to require, it is in fact called for to have all
 fermions arising from fermionization of bosons. To make a
 realistic scenario with the fermions all coming from
 fermionizing some bosons we should still have at least
 some not fermionized bosons and we are driven towards
 that being a gravitational field, that is not fermionized.
 Essentially we reach the spin-charge-families theory by
 one of us (N.S.M.B.) with the detail that the number of fermion
 components and therefore of families get determined from
 what possibilities for fermionization will finally
 turn out to exist. The spin-charge-family theory has long been plagued by
 predicting 4 families rather than the phenomenologically
 more favoured 3. Unfortunately we do not yet understand well
 enough the unphysical negative norm square components
 in the system of bosons that can fermionize in  higher
 dimensions because we have no working high dimensional
 case of fermionization. But suspecting they involve
 gauge fields with complicated unphysical state systems
 the corrections from such states could putatively
 improve the family number prediction.
\end{abstract}


\maketitle

\section{Introduction}

One of the general requirements for quantum field theories 
is microcausality \cite{axiom,axiom2}, the requirement 
of causality, which in its form  
as suggested from tensor product deduction says, that for 
two relative to each other spacelikely placed events 
$x_1$ and $x_2$ in Minkowski space-time a couple 
of quantum field operators ${\cal O}_1$ and ${\cal O}_2$ taken 
at these events will {\bf commute}
\begin{eqnarray}
\{{\cal O}_1(x_1), {\cal O}_2(x_2)\}_{-}&=&0\hbox{  for spacelike 
$x_1 -x_2$}.
\end{eqnarray} 

This is so for the  ${\cal O}_1$ and ${\cal O}_2$ being 
boson fields, but if they are both fermion fields, one 
would have instead to let them {\bf anticommute}
\begin{eqnarray}
\{ {\cal O}_1(x_1), {\cal O}_2(x_2)\}_{+}&=&0 \hbox{  for spacelike 
$x_1 -x_2$}.
\end{eqnarray}
Such {\bf anti}commutation is, however, from the 
tensor product way of arguing for the relation 
completely wrong. We could therefore claim that 
it is not truly allowed to have fermions in the usual 
way, because it leads to a ``crazy'' locality 
axiom. It is one of the purposes of the present 
proceedings article to suggest to investigate the 
consequences of such an attitude, that fermions as 
fundamental particles are not good, but that one should 
rather seek to obtain 
fermions,
not as fundamental, but rather only by fermionization 
of some boson fields instead. But then it becomes very 
important what combinations - what systems - of fermions
can be obtained from appropriate bosonic models. 
For the existence of quite nontrivial restrictions 
on the number of fermions, we can expect to be obtainable 
by fermionization from a system bosons, the 
theorem \cite{Aratyn} 
by Aratyn and one of us (H.B.F.N.) is quite suggestive. In fact this 
theorem tells, that the ratio between the number of 
fermion spin components for all the species (families) counted
together and the corresponding number of boson spin components 
counted together must be $\frac{2^{d_{spatial}}}{2^{d_{spatial}}-1}$.
A priori this theorem seems to enforce that in say the 
experimental number of dimensions, $d_{spatial} = 3$ and 1 time,
the collective number of fermions components must be divisible 
by $2^3 =8$. If we count the components as real fields 
a Weyl fermion has 2*2 = 4 such real components, and thus
the number of Weyl fermions must be divisible by 8/4 =2.

Let us immediately include the remark, that although we shall
below mainly go for the presumably simplest case of 
non-interacting massless bosons - presumably Kalb-Ramond 
fields - being fermionized into also free massless 
fermions, that does NOT mean that we seriously suggest 
Nature to have no interactions. Rather the hope is that 
gravitational degrees of freedom couple in a way specified
alone by the flow of energy and momentum, so that we can 
hope that having a free theory it should be very easy 
and almost unique how to add a gravitational interaction.
Let us say, that by the {\it spin-charge-family} theory by one of us,
the interacting fields are the gravitational ones (vielbeins and spin connections)%
~\cite{IARD,Normas,Normas2} only, but in $d>(1+3)$, fermions 
manifesting in $(1+3)$ as spins and all the observed charges, as well as families,
gravity manifesting all the observed gauge fields as well as the scalar fields, 
explaining higgs and the Yukawa couplings.

In analogy to, how one sometimes says that the electromagnetic 
interaction is added to a system of particles or fields 
with a global charge is ``minimally coupled'', if one 
essentially just replace the derivatives by the 
corresponding covariant ones, we shall imagine that 
our free theory, which has energy and momentum as 
global charges could be made to contain gravity by 
some sort of ``minimal coupling''. To introduce other 
extra interactions than just gravity is, however, expected 
to be much more complicated: Especially higher order 
Kalb-Ramond fields couple naturally to strings and branes,
which in any case would tend to have disappeared in the 
present status of the universe. So effectively to day 
the Kalb-Ramond fields~\cite{KR} should be free except for 
their ``minimal coupling to gravity''. This would mean 
that allowing such a ``later'' rather trivial inclusion 
of gravity, which should be relatively easy, would make 
our at first free model be precisely the since long 
beloved model of one of us, the-spin-charge-family theory~%
\cite{IARD,Normas,Normas2}. Fundamentally  we have thus in our 
picture some series 
of Kalb-Ramond fields together with gravity coupling 
to them in the minimal way. Then we fermionize only this 
series of Kalb-Ramond fields, but keep bosonic the gravitational
field, which probably cannot be fermionized even, if we 
wanted to. The resulting theory thus becomes precisely of 
similar type as the one by one of us, the spin-charge-family theory.

Now, however, the Kalb-Ramond fields are plagued by a lot 
of gauge symmetry and ``unphysical'' degrees of freedom,
some of which even show up with even negative norm squared
inner products.
In principle these unphysical degrees of freedom must 
also somehow be treated in the fermionization procedure.
Especially, if we want to use our theorem of counting degrees
of freedom under bosonization~\cite{Aratyn}, we should 
have such a theorem allowed to be used also when the 
``unphysical'' d.o.f. are present.
  
In fact it is the main  
new point in the present article,
that we put forward a slightly more complicated 
Aratyn-Nielsen-theorem - an extended Aratyn-Nielsen theorem -, 
allowing for theories with negative norm
squared normalizations.      

It is the true motivation of the present work, that once 
when we shall find some genuinely working case(s) of 
theories that bosonize/fermionize into each other in 
high dimensions, they will almost certainly turn out 
to involve  gauge theories on the bosonic side. That is to say 
it will be combinations of various Kalb-Ramond fields~\cite{KR} 
(among which we can formally count also electromagnetic 
fields and even a scalar field), and such Kalb-Ramond fields 
often have lots of negative norm square components. 
Thus once we know what is the boson theory that can be 
fermionized we need an extended Aratyn-Nielsen theorem
to calculate the correct number of fermion components matching 
the fermionization correspondance. 
Well really, if we know it well, we can just read off 
how many fermion components there are.
 It is namely this 
number of fermion components, that translates into the number 
of families, on which they are to be distributed. It means 
that knowing the detailed form of the boson system 
and the rule - the extended Aratyn-Nielsen theorem -
for translating the number of boson components into the 
number of fermion components is crucial for obtaining 
the correct number of families. Will so to speak 
the number of fermion-families remain 4 as claimed 
by one of us in her model, which has reminiscent 
of being a fermionization, or will it be corrected 
somehow from the true bosonization requirement including the 
negative norm square components for the bosons? 
The reliability of the model would of course  - according 
to the judgement of one of us (H.B.F.N.) - be much bigger, 
if it turned out that the true prediction were 3 families 
rather than the 4 as usually claimed, except, of course,
if the fourth family, predicted by the spin-charge-family theory,
 will be measured.
  
With the old Aratyn-Nielsen theorem (the unextended version)
it does crudely not look promising to get the number 3 rather
than 4 as H.B.F.N  would hope phenomenologically in as far this 
version implies that the number of fermion components is 
divisible by a
rather high power of the number 2. Such a number-theoretic 
property of the number of families seems a priori to 
favor 4 much over 3.

The works of major importance for the present 
talk are:
\begin{itemize}
\item{\color{blue} Aratyn \& Nielsen} We made a theorem ~\cite{Aratyn} 
about the ratio of the number of bosons needed to represent 
a number of fermions based on  statistical mechanics in the free 
case, under the provision that a bosonization 
exists. 
\item{\color{blue} Kovner \& Kurzepa} They\cite{KK,KK2} present an 
explicit bosonization of two complex fermion fields 
in 2+1 dimensions being equivalent to $QED_3$ meaning 
2+1 dimensional quantum electrodynamics.
\item{\color{blue} Manko\v c-Bor\v stnik}~\cite{IARD,Normas,Normas2} 
The spin-charge-family unification 
theory explains the number of families from the 
number of fermion components appeared in  this theory.
\end{itemize}

In the next section \ref{hope} we put forward the main 
hope or point of view of our application of bosonization 
to make prediction of the number of families. In section \ref{guiding} we give a loose argument for
what we think should our picture for nature to cope with 
the investigations in the present article.
Then we shall 
in section \ref{Review} and \ref{AratynN} 
review both Kalb Ramond fields and and our old Aratyn-Nielsen 
theorem about the number fermion components needed to 
make an equivalent theory with a number of boson components. 
In section \ref{Time} we look at the problem, that the 
components of a Kalb-Ramond field with an index being $0$
are on the one hand to be a conjugate momentum to the 
other components and on the the other hand, if we use 
Lorentz invariance, have to lead to states with 
negative norm square. The latter is of course simply 
a reflection of the signature of the Minkowski metric tensor.
It is for the application on such negative norm square 
components - the components with an index $0$ - that 
our extension of our Aratyn-Nielsen theorem to negative 
norm square components become relevant.   

In  section \ref{Kovner} we review the work by 
Kovner and Kurzepa\cite{KK}, who proposed a concrete bosonization 
including explicit expressions for the fermion fields 
in terms of the boson fields - actually simply electrodynamics -
in the case of 2 space dimensions and one time, 1+ 2. Next
in \ref{Match} we seek to check our Aratyn-Nielsen theorem
on this special case of 1+2 both by counting the 
particle species including spin states~\ref{Particles} and 
by counting the fields~\ref{Fields}. 

Towards the end, section~\ref{New}, we seek to reduce away 
some of the degrees 
of freedom from the Kovner and Kurzepa model to obtain 
a reduced case with fewer particles on which we - if it 
is also a case of bosonization - would  be again able to 
check our counting theorem (Aratyn-Nielsen).

\section{Hope}
\label{hope}
{\bf Use of Bosonization/Fermionization Justifying Number of
Families}

The governing philosophy and motivation for the 
present study is:

\begin{itemize}
\item Fermions do NOT exist fundamentally (because they do not have 
proper  causal/local property) .
\item Some boson degrees of freedom are rewritten by 
bosonization (better fermionization)  to fermionic ones,
which then make up the fermions in the world, we see.
(but some other boson degrees of freedom, hopefully gravity,
are not bosonized). 
\item We work here only with an at first free theory 
- for our presentation, it might be best if only 
bosonization worked for FREE theories in higher dimensions -
i.e. free bosons can be rewritten as free fermions.
\item We though suggest - hope- that exterior to
both bosons and/or  fermions, we can add a GRAVITATIONAL
theory. So fundamentally: gravity with  
matter bosons. It gets rewritten to fermions in a 
gravitational field, just similar to the 
theory~\cite{IARD,Normas,Normas2} of one of us called spin-charge-family  
unification theory. 
 \end{itemize}

Let us be more specific about the dream or hope behind 
the present project: 

By using say ideas from the 
below discussed paper by Kovner and Kurzepa~\cite{KK} or by our 
own earlier article in last years Bled Proceedings 
about bosonization, we hope to find at 
least a case of fermionizing some series of Kalb-Ramond
fields (i.e. Boson fields) - and electrodynamics is of 
course considered here a special Kalb-Ramond one - into 
some system of fermions. Presumably it is easiest - and 
perhaps only possible - for free theories or only theories 
interacting in a very special way. We therefore are most 
eagerly going for such a free and even massless case. 

But now if indeed we can find such a case, or if exists, 
then it is very likely that we can extend it to interact 
with gravity in a minimal way. 
In fact we all the time require our hoped for fermionization 
cases to have the same energy and momentum for the 
bosonic and the fermionic theories that shall be equivalent.
Thus the fermionization procedure, if it exist at all, is
compatible with energy and momentum. 

If we therefore let our boson-theory interact with gravity,
that couples to the energy and momentum - specifically to 
the energy momentum tensor $T_{\mu\nu}(x)$ - we have some hope 
that this coupling of the boson fields to gravity will 
simply transfer to a coupling of the fermionized 
theory, too.

As procedure we might have in mind writing the free massless
fermionization procedure in arbitrary coordinates. That should 
of course be possible, but although the theory would 
now look as a gravitational theory, it would only have been
derived for the case of the gravitational fields 
having zero curvature, i.e. for the Riemann tensor being 
zero all over. However, if the fermionization procedure 
could be described by a local expression for the Kalb-Ramond 
fields - or other boson fields - expressed in terms of 
fermion currents or the like, then the correspondence 
would in that formulation be local and lead to the 
energy momentum tensor being also related in such simple 
local way. I.e. we would have in this speculation 
\begin{equation}
T_{\mu\nu}(x)|_{boson} = T_{\mu\nu}(x)|_{fermion}.
\end{equation}   
Here of course the two energy momentum tensors are the ones
in respectively the fermion and the boson theory being 
equivalent by the dream for fermionization.

It is further our hope for further calculation that 
we may argue that in general it is very difficult to 
have interaction with Kalb-Ramond fields except for 
\begin{itemize}
\item The appropriate branes,
\item Some general gauge-theory coupling to the 
charges (think of global ones) conserved by the 
Kalb-Ramond- theory in question. But since the always 
conserved global charges are the energy and momentum 
this suggests the coupling to gravitational field.  
\end{itemize}

We thus want to say that this starting form fundamental 
Klab-Ramond fields supposedly difficult to make interact
points towards a theory at the end with gravity as the 
only interaction. Gravity namely is suggested to 
be hard to exclude as possibility even for otherwise 
difficult to make interact Kalb-Ramond fields.

If we manage to fermionize the Kalb-Ramond fields as just 
suggested, we therefore tend to end up with the spin-charge-family 
unification model of one of us in the sense that we get 
ONLY gravity interaction, and otherwise a free theory.

But it shall of course be understood here that we only 
fermionize some of the boson fields in as far as we 
leave the assumed fundamental gravity field non fermionized.

\section{Guiding and Motivation}
\label{guiding}
The reader might ask why we choose - and suppose Nature 
to choose - these Kalb-Ramond-type fields which are to be 
explained a bit more below in section (\ref{AratynN}).
Let us therefore put forward a few wish-thinking arguments
for our bosonic fundamental model:

\begin{itemize}
\item We have no way to make fermionization/bosonization 
conserving angular-momentum truly (at the same time keeping
the spin statistics theorem): The bosons namely 
necessarily can only produce Fock space states with 
integer angular momentum, but the fermion  sates should for 
an odd number of fermions in the the Fock state have 
half integer angular momentum. So clearly 
fermionization/bosonization conserving angular momentum is 
{\em impossible}!
\item The trick to overcome this  angular-momentum-problem
is to {\em reinterprete a spin 1/2 index on the fermions
as a family index instead}. That is to say we accept at first
that the fermions come out of the fermionization with 
bosonic integer spin index combination, and then seek to 
reinterprete part of the spin polarisation information 
as instead being a family information.

\item In fact we shall be inspired by the spin-charge-family 
unification model to go for that the fermions come out 
from the fermionization at first with {\em two} spinor indices,
so that they have indeed formally at this tage {\em integer} 
spin. Then we make the interpretation that one of these 
spinor indices is indeed a {\em family index}. That of course
means, that we let one of the two indices be taken as 
a scalar index i.e.  being not transformed under Lorentz 
transformations. 
\item So we decide to go for a system of fermions 
at the ``first interpretation'' being a two-spinor-indexed
field. But now such a field $B_{\alpha\beta}$, where $\alpha$ 
and $\beta$ are the spinor indices, is indeed a Clifford 
algebra
element, or we could say a Dirac matrix (or a Weyl matrix 
only if we use only the Weyl components). In any case 
we can expand it on antisymmetrized products of gamma-matrices:
\begin{multline}
B_{\alpha\beta} =
 \bigl( a {\bf 1} + a_{\mu}\gamma^{\mu} +\dots
+ a_{\mu\nu...\rho}\gamma^{\mu}\gamma^{\nu} \cdots\gamma^{\rho}+\dots \\
 +a_{0,1,...,(d-1)}\gamma^0\gamma^1\cdots 
\gamma^{(d-1)} \bigr )_{\alpha\beta},
\end{multline} 
and thus the boson fields suggested to by fermionization 
leading to such fermion fields should be a series of 
antisymmetric tensor fields of all the different orders
from the scalar $a$ and the d-vector $a_{\mu}$ all the way 
up to the maximal antisymmetric order tensor $a_{0,1,...,(d-1)}$.

\item With random coefficients on a Lagrange-density 
expansion for a theory with boson fields, which have 
d-vectorial indices one unavoidably loose the bottom 
in the Hamiltonian as one can see from e.g. just 
a term like 
\begin{eqnarray}
c*\left ( \partial_{\mu}\cdots\partial_{\nu} a_{\rho...\tau}\right )^2. 
\end{eqnarray}
Think for instance on the terms for 
which the series of the derivative indices are 
spatial so that we have to do with a potential 
energy term. If the coefficient $c$ is adjusted to let
the contribution with the indices on $a_{\rho...\tau}$ being 
spatial
to the Hamiltonian be positive, then the contributions 
with a $0$ among these indices will from Lorentz 
invariance have to be of the wrong sign. So it is 
at best exceedingly hard to organize a positive definite 
Hamiltonian density. 

Consider only the free part - meaning bilinear part in the 
field $a_{\rho...\tau}$ - in the Lagrangian. For simplicity let 
us consider 
the situation of a field $ a_{\rho...\tau}$ being 
constant as function of the time coordinate $x^0$,
and that the number of derivatives acting on the field 
is so low that some of the indices- say $\rho$ - on 
the $a_{\rho...\tau}$
has to be contracted with another one or the same index
on this field in order to cope with Lorentz invariance.
Then if this (sum of) squares of the field in some combination 
shall get a for the hamiltonian positive contribution 
from a spatial  value of the index $\rho$, it will get 
the opposite sign for $\rho =0$. So it looks that we cannot
avoid the Hamiltonian having both signs for a 
``free term'' in the Lagrangian, unless all the indices 
on $a_{\rho...\tau}$ are in the term contracted with derivatives.
But with the  antisymmetry this would be zero for more than 
one index on $a_{\rho...\tau}$. So indeed it seems that 
unless one gets the fields restricted in some way, so that 
these fields or their conjugate variables are somehow 
not allowed to take independent values, then the Hamiltonian
will loose its bottom and (presumably infinite) negative
energy values will be allowed.
\item We are thus driven towards theories with constraints!
\item Such constraints are typically obtained by means 
of some gauge symmetry, and thus we are driven towards 
theories with gauge symmetry, if we want to uphold
a positive definite Hamiltonian for the by the constraints 
allowed states of the field and its conjugate momenta.
\item The obvious candidate for such a gauge theory with 
antisymmetrised tensor fields is of course the Kalb-Ramond
fields. (Personally we suspect, that we can show that 
ONLY Kalb-Ramond-fields will solve this problem 
of positivity by providing enough constraints.)
\item Thus it seems that it is very hard to hope 
for our to be used fermionization unless we make use 
of presumably a whole series of Kalb-Ramond fields!           
\end{itemize}
\section{Review}
\label{Review}
In theoretical condensed matter physics and particle physics, 
Bosonization/fermio\-nization  is a mathematical procedure by which a system of 
possibly interacting fermions in (1+1) dimensions can be 
transformed to a system of massless, non-interacting bosons. 
In the present article we shall dream about extending such 
bosonization to higher dimensions, and we shall be most 
interested in the case when even the fermions do not 
interact. 
 The method of bosonization was conceived independently 
by particle physicists Sidney Coleman and Stanley Mandelstam; 
and condensed matter physicists Daniel Mattis and Alan Luther 
in 1975. \cite{bosonization}
The progress to higher dimensions has been less developed
\cite{higher} than 
the 1+1 dimensional case, but there has been some works 
also on higher dimensions. especially we shall below 
review a bit a work\cite{KK} by Kovner and Kurzepa for the next 
to simplest case, namely 2+1 dimensions. There has also 
been developments based on Chern-Simon type action\cite{higher}, 
but we 
suspect that the type of bosonization we are hoping for 
in the present article should rather be of the Kovner Kurzepa
type than of the Chern-Simon one, although we have difficulty
in explaining rationally why we believe so.

When we have such transformation and thus two equivalent
theories, one with fermions and one with bosons, one will
of course expect that the number of degrees of freedom 
should in some way be the same for the boson and for the 
fermion theory. Otherwise of course they could not
be equivalent. In the most studied case of 1+1 dimensions
it has turned out in the cases known that there are 
two fermion components per boson component. This ratio 
is in accord with the theorem by Aratyn and one of us
\cite{Aratyn}
- which we call the Aratyn-Nielsen theorem - in as far as 
this theorem predicts the ratio to be 
$\frac{2^{d_{spatial}}}{2^{d_{spatial}}-1}$ where $d_{spatial}$ is
the dimension of space ( not including time) so that we 
talk about the dimension $d_{spatial} +1$. In fact of course 
for the case 1+1 we have thus $d_{spatial} =1$ and the fraction
predicted becomes $\frac{2^1}{2^1-1} =2 $ times as many 
fermion components as boson components. This is really 
assuming that a ``component'' corresponds to a polarization 
state of a particle. What we - one of us and Aratyn -
really derived was that for a theory with massless interacting
there ahd to be the mentioned ratio between the 
number of polarization states for the fermion(s) relative
to that for the bosons. It were namely the contributions 
of such polarization states to the average energy in a Boltzmann 
distribution calculation that was used to derive the theorem.
Although derived for this non-interacting massless case there 
could be reasons to believe that by taking a couple of limits 
in an imagined case of interacting and perhaps massive 
bosonization it could be argued, that the theorem of 
ours would have to hold anyway. For instance going to a very 
small distance scale approximation an approximately
massless theory would arrive and the theorem should be 
 applicable even if there is a mass.
Since we are concerned in this 
theorem really with a counting of degrees of freedom 
a very general validity is in fact, what would be expected.
As already said, we are, however, in the present article 
more concentrating on the generalization to include 
some unphysical degrees of freedom with possibly 
wrong signature,

\section{AratynN}
\label{AratynN}
{\bf \large Aratyn-Nielsen Theorem for massless free 
Bosonization}

{\bf If } there exist two free massless quantum field 
theories respectively with Boson, and Fermion particles and
they are equivalent w.r.t. to the number of states of 
given momenta and energies, {\bf then} the two theories
must have the same average energy densities for a given 
temperature $T$, or simply same average energies, if 
we take them with the same infrared cut off(a quantisation 
volume $V$):
\begin{eqnarray}
<U_{boson}> &=& <U_{fermion}>
\hbox{where}\\
<U_{boson}> &=&
\sum_{\vec{p}}\frac{E(\vec{p})}{1-\exp{(E(\vec{p})/T)}}\\
<U_{fermion}> &=&
\sum_{\vec{p}}\frac{E(\vec{p})}{1+\exp{(E(\vec{p})/T)}}.\\
\end{eqnarray}  
Here $\vec{p}$ runs through the by the infrared cut off 
allowed momentum eigenstates, and $E(\vec{p})$ are the 
corresponding single particle energies.
Of course the single particle energy for a mass-less theory is 
\begin{equation}
E(\vec{p}) = |\vec{p}|,
\end{equation}
when c=1, and in $d_{spatial} $ dimensions and with an infrared
cut off spatial volume $V$ the sum gets replaced in the 
continuum limit by the integral
\begin{equation}
\sum_{\vec{p}} ... \rightarrow 
\int \sum_{components} ... \frac{V}{(2\pi)^{d_{spatial}}},
\end{equation} 
where $\sum_{components}...$ stands for the sum over the 
different polarization components of the particles in question.
So effectively in the simplest case of all the particles having
the same ``spin''/the same set of components we have 
the replacement
\begin{equation}
\sum_{components} ... \rightarrow N_{families} * N_c...
\end{equation} 
where $N_c$ is the number of components for each particle 
and $N_{families} $ is the number of families.
{\bf Some formulas for deriving Aratyn-Nielsen}
\begin{eqnarray}
<U_{boson}> &=&
\sum_{\vec{p}}\frac{E(\vec{p})}{1-\exp{(E(\vec{p})/T)}}\\
&=&``N_{families} *N_c '' *V/(2\pi)^{d_{spatial}}*\\
&& \int O(d_{spatial} )
|\vec{p}|^{d_{spatial}} E(\vec{p})\sum_{n=0, 1, ...}\exp{(nE(\vec{p})
/T} d|\vec{p}|\nonumber\\
\end{eqnarray}


{\bf Simple Aratyn-Nielsen Relation}
For a given temperature must the average energies of 
respectively the boson and the with it equivalent fermion 
theories 

{\bf \large Our Realization Suggestion}

\begin{itemize}
\item{Fermions}

For the fermions we shall use the needed number of 
say Weyl fermions, i.e. we must adjust the number 
of families hoping that we get an integer number.

\item{Bosons}

For the bosons we let the number $2^{d_{spatial}} -1$ suggest 
that we take a series of all Kalb-Ramond fields, one combination
of fields for each value of the number $p$ of indices on the 
``potential field''  $A_{ab...k}$ (where then there 
are just $p$ symbols in the chain $ab...k$). At first we take 
these symbols $ a,b, ...,k$ to be only spatial coordinate
numbers.

\end{itemize}

{\bf \large Free Kalb-Ramond }
A Kalb-Ramond field\cite{KR} with p indices on the 
``potential'' and p+1 indices on the strength 
\begin{equation}
F_{\mu\nu\rho...\tau}(x)= \partial_{[\mu}A_{\nu\rho...\tau]}(x),
\end{equation}
where $[...]$ means antisymmetrizing, and the ``potential''
$A_{\nu\rho ...\tau}$ is antisymmetric in its $p$ indices 
$\nu\rho ...\tau$, is defined to have an action invariant
under the gauge transformation:
\begin{equation}
A_{\nu\rho...\tau}(x) \rightarrow A_{\nu\rho...\tau}(x) 
+\partial_{[\nu}\lambda_{\rho...\tau]}(x)
\end{equation}
for any arbitrary antisymmetric gauge function 
$\lambda_{\rho...\tau}(x)$ with $p-1$ indices.

{\bf Free Kalb-Ramond Action:}

Note that the strength $F_{\mu\nu\rho ...\tau}=
\partial_{[\mu}A_{\nu\rho...\tau]}$ is gauge invariant, and that
thus we could have a gauge invariant Lagrangian density
as a square of this field strength
\begin{eqnarray}
{\cal L}(x) &=& F_{\mu\nu...\tau}F_{\mu'\nu'...\tau'} g^{\mu\mu'}*
g^{\nu\nu'}*...*g^{\tau\tau'}.
\end{eqnarray} 
Then the conjugate momentum of the potential becomes(formally):
\begin{eqnarray}
\Pi_{\nu\rho...\tau}=\Pi_{A_{\nu\mu...\tau}}&=&\frac{\partial{\cal L}}
{\partial{(\partial_0 A_{\nu\rho...\tau})}}\nonumber\\
&=&F_{0\nu\rho...\tau}.
\end{eqnarray}

{\bf A Lorentz gauge choice:}
\begin{eqnarray}
\partial_{\mu}A{\nu\rho...\tau} g^{\mu\nu}&=&0,
\end{eqnarray}
allows to write the Lagrange density instead as
\begin{eqnarray}
{\cal L}_{modified}(x)&=& 1/2 *\partial_{\mu}A_{\mu\nu...\tau}
\partial_{\mu'}A_{\mu'\nu'...\tau'}
*g^{\mu\mu'}g^{\nu\nu'}\cdots g^{\tau\tau'},  
\end{eqnarray} 
which leads to the very simple equations of motion letting 
each component of the ``potential'' $A_{\nu\rho...\tau}$
independently obey the Dalambertian equation of motion
\begin{equation}
g^{\mu\mu'}\partial_{\mu}\partial_{\mu'}A_{\nu\rho ...\tau}=0.
\end{equation}
{\bf Lorentz Invariance Requires Indefinite Inner Product!:}

Lorentz invariant norm square for the states generated 
by the creation operators $a_{\nu\rho...\tau}^{\dagger}(p)$, 
i.e. $a_{\nu\rho...\tau}^{\dagger}(p)|0>$, must have different 
sign of the norm square depending on whether there is 
an even (i.e. no) $0$'s among the indices or whether there 
is an odd number (i.e. 1). A priori we are tempted to 
take 
\begin{eqnarray}
<0|a_{\nu\rho...\tau}(p)a_{\nu\rho...\tau}^{\dagger}(p)|0>& >&0 
\hbox{ for no $0$ among the indices,}\nonumber \\
 <0|a_{\nu\rho...\tau}(p)a_{\nu\rho...\tau}^{\dagger}(p)|0>& <&0 
\hbox{ for one $0$ among the indices,}\nonumber \\
\end{eqnarray}
  

\section{Time-index}
\label{Time}
{\bf Problem with Components with the time index $0$:}

But full Kalb-Ramond fields require also components 
a $0$ among the indices.(This is the main new thing 
 in the present article
to 
treat this problem of the components with one 
$0$ among the indices.)

Remember about these components with a $0$ index:
\begin{itemize}
\item Using a usual Minkowskian metric tensor $g^{\mu\nu}$ in 
constructing an inner product between Kalb-Ramond fields, 
say 
\begin{equation}
g^{\mu\nu}g^{\rho\sigma} \cdots g^{\tau\kappa} A_{\mu\rho...\tau}
 ( \hbox{potentially an $\partial^0$} ) A_{\nu\sigma ...\kappa},
\end{equation} 
we get the opposite signature (=sign of the square norm) 
depending on whether there is a $0$ or not!

{\bf This means that if particles produced by the components
without the $0$ index have normal positive norm square, then 
those produced by the ones with 
the $0$ have negative norm-square!}
\end{itemize}
{\bf Good Luck We Removed the Kalb-Ramond $A$ with $p=0$ 
Indices!}
We could namely not have replaced on $A$ one among its indices 
by a $0$ because it has no indices. So we would not have 
known what to do for the 
fields $A$ with $0$ indices. 

We correspondingly also have to leave out the Kalb-Ramond-field
with $p=d_{spatial} +1$ indices, because for that there would 
be no components without an index $0$. 

For the unexceptional index numbers $ p= 1,2,..., d_{spatial}$ 
there are some components both with and without the $0$.

For the two exceptions $p = 0 \hbox{ and } d=d_{spatial} +1$
we have chosen not to have a Kalb-Ramond-field in our 
scheme, using it to get the $-1$ in the from 
Aratyn-Nielsen required $2^{d_{spatial}}-1$.  

{\bf Simplest (Naive) Norm Square Assignment}

Note that for each Kalb-Ramond-field we can choose an overall
extra sign on the inner product, because we simply can 
define the overall inner product with an extra minus sign,
if we so choose. But the simplest choice is to just let 
the particles corresponding to fields with {\color{blue}
only spatial 
indices} (i.e. all $p$ indices different from $0$) to 
have positive {\color{blue}norm square}, while then 
{\color{red} those with one $0$} have {\color{red} negative 
norm square}.  

{\bf This simple rule would lead to equally many 
components/particles with positive as with negative norm 
square, so that dreaming about imposing a constraint that 
removes equally many negative and positive norm square at a 
time would leave us with nothing.}

{\bf Numbers of Components with and without $0$.}
An of course totally antisymmetric field $A_{\mu\nu...\tau}$ with
$p$ indices has
\begin{eqnarray}
\hbox{\# components}_{KR\; p \hbox{indices}}&=&{d \choose p} 
={ d_{spatial} +1 \choose p}\nonumber\\
\hbox{\# no $0$ components}_{KR\; p \hbox{indices}}&=& {d_{spatial} 
\choose p} = {d-1 \choose p}\nonumber\\
\hbox{\# cmps. with $0$ \& p-1 non-$0$}_{KR \; p 
\hbox{indices}}&=&{d_{spatial} \choose p-1}= {d-1\choose p-1}.
\nonumber
\end{eqnarray} 
and so one must have as is easily checked
\begin{eqnarray}
{d \choose p}& =&{\color{red} {d-1 \choose p}}
+ {\color{blue} {d-1 \choose p-1}}\nonumber\\
\hbox{corresponding to}&&\nonumber\\
\hbox{``All components''} &=& {\color{red} \hbox{``Without $0$''}}+
{\color{blue}\hbox{``With $0$''}} \nonumber
\end{eqnarray}

{\bf Using ONLY the Components WITHOUT $0$ would fit $2^{d_{spatial}} $ Nicely !}
Having decided to leave out the number of indices $p$ values
$ p=0$  and $ p=d$ the number of components {\bf without} 
any component indices being $0$ just makes up 
\begin{eqnarray}
\hbox{\# without $0$ }_{\hbox{for all $p=1,2,...,d-1$}}&=&
\sum_{p=1,2,...,d-1}{d-1 \choose p}
=2^{d-1} -1\nonumber
\end{eqnarray}
so these ``only with spatial indices components'' could
elegantly correspond to $2^{d-1}$ = $2^{d_{spatial}}$ fermion 
components.

{\bf But problem: Kalb- Ramond fields need also the component
with an index being $0$.}  

{\bf Using ONLY the Components WITH $0$ could also 
fit $2^{d_{spatial}} $ Nicely !}
Having decided to leave out the number of indices $p$ values
$ p=0$  and $ p=d$ the number of components {\bf with} 
the  $0$ just makes up 
\begin{eqnarray}
\hbox{\# with $0$ }_{\hbox{for all $p=1,2,...,d-1$}}&=&
\sum_{p=1,2,...,d-1}{d-1 \choose p-1}\nonumber\\
&=&2^{d-1} -1\nonumber
\end{eqnarray}
also, so these ``only with $0$  index components'' could
elegantly correspond to $2^{d-1}$ = $2^{d_{spatial}}$ fermion 
components, also!

{\bf But problem: Kalb- Ramond fields need also the components
without an index being $0$, and these with $0$ usually 
come with wrong norm square.}  

{\bf The Trick Suggested is to use for Some KR-fields Opposite
Hilbert Norm Square}

In other words we shall look along the chain of all 
the allowed $p$-values $p = 1, 2, ..., d-1;$ and for 
each of these $p$-values we can choose whether 
\begin{itemize}
\item{\color{blue} Normal: } The states associated with the 
polarization components {\bf without} the $0$ among the 
indices shall be of {\bf positive norm square}, as usual, 
and then from Lorentz invariance essentially the ones 
{\bf with the $0$ shall have negative norm square}, or

\item{\color{blue} Opposite} The states {\bf with $0$ shall
have positive } norm square, while the components 
{\bf without $0$  negative } norma square.    
\end{itemize}  
{\color{red} Our proposal: Choose so that we get the 
largest number of positive norm square components.} 
{\bf How to get Maximal Number of Positive over Negative Norm 
Square Single Boson States}

For each value of $p$ (=the number of indices on the 
Kalb Ramond ``potential'') $p = 1,2,..., d_{space}$ decided 
to be used in the bosonization ansatz a priori, we investigate 
whether the number of (independent) components {\bf with} or 
{\bf without} a $0$ is the bigger:
\begin{eqnarray}  
\hbox{\# no $0$ components}_{KR\; p \hbox{indices}}&=& {d_{spatial} 
\choose p} = {d-1 \choose p}\nonumber\\
\hbox{\# cmps. with $0$ \& p-1 non-$0$}_{KR \; p 
\hbox{indices}}&=&{d_{spatial} \choose p-1}= {d-1\choose p-1}.
\nonumber
\end{eqnarray}
So if there are {\bf most components without $0$},
i.e. if ${ d_{spatial}\choose p-1} < {d_{spatial} \choose p}$,
then we give the particle states  corresponding to the
{\bf without $0$} ``potentials'' have {\bf positive norm square}.
And opposite if  ${ d_{spatial}\choose p-1} > 
{d_{spatial} \choose p}$.

But if there are {\bf most components with $0$},
i.e. if ${ d_{spatial}\choose p-1} > {d_{spatial} \choose p}$,
then we give the particle states  corresponding to the
{\bf with $0$} ``potentials'' have {\bf positive norm square}.

{\bf To Maximize Positive Norm Square we Choose:}
\begin{itemize}
\item When $ p< \frac{d}{2}$, choose {\bf without $0$ positive
                norm} squared, while ``with $0$'' negative;
\item but when  $p>  \frac{d}{2}$, choose {\bf with $0$ positive norm} squared, while ``without $0$'' negative;
\end{itemize}
For e.g. $p<d/2$ the excess of positive norm square 
``components '' over the negative norm ones becomes:
\begin{eqnarray}
&&{ d_{space} \choose p} - {d_{space} \choose p-1}={ d_{space} \choose p}(1 - \frac{p}{d_{space}-p+1}) \nonumber \\
&&=\frac{d_{space}!(d_{space}+1-2p)}{(d_{space}-p+1)!p!}
= \frac{(d-1)!(d-2p)}{(d-p)!p!},
\end{eqnarray}

However, for $p>d/2$ the excess is
\begin{eqnarray}
&&{ d_{space} \choose p-1} - {d_{space} \choose p}=
{ d_{space} \choose p-1}(1 - \frac{d_{space}-p+1}{p}) \nonumber \\
&&=\frac{d_{space}!(2p -d_{space}-1)}{(d_{space}-p+1)!p!}
= \frac{(d-1)!(2p-d)}{(d-p)!p!},
\end{eqnarray} 

{\bf Adding up Positive Norm Square over Negative Excess:}

The sums over $p$ `` telescopes'' from each of the two 
cases of p bigger or smaller than d/2, and gives by symmetry 
the same excess of positive over negative norm square states,
namely for each for say $d$ even (i.e. $d_{space}$ odd)
\begin{eqnarray}
{d-1 \choose d/2 -1} -1&=&\frac{(d-1)!}{(d/2 -1)!(d/2+1)!}-1,
\end{eqnarray}
where we used that the middle value $p=d/2$ contribution 
vanishes. Including as we shall both ``sides'' smaller than 
d/2 and also bigger than d/2 we get the double of this.

{\bf Example Excesses States for even $d$ for Bosons}
\begin{eqnarray}
Excess(d=2) &=&2({2-1 \choose 2/2-1} -1) =0\nonumber \\
Excess(d=4)&=& 2({4-1\choose 4/2 -1} -1)=2 \nonumber \\
Excess(d=6)&=& 2({6-1\choose 6/3 -1} -1)= 18 \nonumber \\
Excess(d=14)&=& 2({14-1\choose 14/2 -1}-1)  
\end{eqnarray}
{\bf Contribution from a Negative Norm square Component}

One shall count the Hilbert space states with the negative 
norm square into the Boltzmann weighted averaging with 
a minus extra.

This extra minus for a negative norm square boson functions 
accidentally just like the 
fermi-statistics versus bose statistics. And thus e.g. a small 
p timelike polarization contributes to the average energy 
just like a fermion, though with an over all minus sign. 
\section{Extension of Our Theorem on Counting}
\label{Extension}
It is a major purpose of the present talk to present 
an extension of the Aratyn-Nielsen theorem\cite{Aratyn} on 
the numbers 
of bosons versus fermions in a bosonization to include 
the just above discussed negative norm square states 
associated with the Kalb-Ramond components having an index 
0. Since such states obtaining at first negative norm squares
are seemingly enforced by Lorentz invariance, it seems to be 
important to extend our Aratyn-Nielsen theorem to the case, 
where some of the components of the fields are quantized with 
a negative norm square. 

We take such a negative norm square mode to mean, that 
whenever there in a Fock space state is an {\em odd} number
of particles with the component in question, then such 
a Fock-space basis vector is in the ``Hilbert norm'' given
a negative norm square. Of course that means that strictly
speaking our Fock space is no longer a genuine 
Hilbert space, but rather just an (infinite dimensional) 
space with an {\em indefinite}  inner product, $|$, giving 
the inner product between two Focks,  $|a>$ and $|b>$ say,
as $<b|a>$. But now the point is just 
that {\em we have no sign restriction on $<a|a>$; it can easily 
be negative.}

The in usual Hilbert spaces used expansion on an 
orthonormal basis 
\begin{eqnarray}
{\bf 1 }&=& \sum_a |a><a| \hbox{ (usual)},
\end{eqnarray}
cannot now be applied. Now we rather have to use
\begin{eqnarray}
{\bf 1} &=& \sum_a (-1)^{N_{neg}(a)}|a><a| 
\hbox{ (with negative norm square also)},
\end{eqnarray}
where $N_{neg}(a)$ denotes the number of particles in 
the various negative norm square single particle states 
together. If for instance a basis state $|a>$  for the Fock 
space has 3 particle in the states with 0 index all together
(and we have used the choice of letting the components 
with a $0$-index be the ones with negative norm, rather than 
the more complicated possibilities discussed above),
$N_{neg}(a)=3$ and thus such a state would come with 
a minus sign in the expansion of the unit operator ${\bf 1}$. 

Let us now calculate the average energy for a system 
described by a Fock space with only one single particle 
state present, so it really is the system with only 
one single particle state, that may be filled or empty 
according the rule for it being bosonic or fermionic 
and having negative or positive norm square. For this purpose
we have to think about how one shall define the concept 
of a trace - which goes into the average procedure 
to provide us with such a an average of the energy, and we 
claim that we must indeed in the case with negative norm square 
states take the trace definition:
\begin{eqnarray}
Tr ( {\bf O})&=& \sum_a (-1)^{N_{neg}(a)}<a|{\bf O}|a>.
\label{Tr}
\end{eqnarray}
With this definition we easily check some 
usual rule for traces:
\begin{eqnarray}
Tr ({\bf O}{\bf P} ) &=& \sum_a (-1)^{N_{neg}(a)}
<a|{\bf O}{\bf P}|a>\\
&=&   \sum_a \sum_b (-1)^{N_{neg}(a)}
<a|{\bf O}|b><b|(-1)^{N_{neg}(b)}{\bf P}|a>\\
&=& Tr({\bf P}{\bf O}).
\end{eqnarray}

Using this definition of the trace $Tr$ we can then 
put in the quite analogous way to the usual case
for Boltzmann distribution in quantum mechanics
\begin{eqnarray}
<E> &=& \frac{Tr(\exp(-H/T) H)}{Tr(\exp(-H/T))}, 
\end{eqnarray}
where the Boltzmann-Constant $k$ has been absorbed into the 
temperature $T$, and where now we use in the case of negative 
norm square the expression (\ref{Tr}). Let us enumerate the 
single particle states with the letter $n$ and denote the 
single particle energy of the state $n$ as $E_n$. Then the 
free Hamiltonian $H$ is given by means of the number operators
\begin{equation}
N_n = a_n^{\dagger}a_n
\end{equation}
as 
\begin{equation}
H= \sum_n E_nN_n = \sum_nE_na_n^{\dagger}a_n, 
\end{equation}   
and we immediately see that 
\begin{eqnarray}
&&<E_nN_n>|_{boson\, pos.} =
\frac{\sum_{N_n=0,1,...}E_nN_n\exp(-E_nN_n/T)}{\sum_{N_n=0,1,2,...}
\exp(-E_nN_n/T)}\nonumber\\
&&=  \frac{-\frac{d\left( \frac{1}{1-\exp(-E_n/T)} \right )}{
d(1/T)}}{\frac{1}{1-\exp(-E_n/T)}}
= \frac{E_n}{\exp(E_n/T)-1} \hbox{(boson; pos. norm sq.)}\\
&& <E_nN_n>|_{boson\, neg.} =
\frac{\sum_{N_n=0,1,...}(-1)^{N_n}E_nN_n\exp(-E_nN_n/T)}{\sum_{N_n}
(-1)^{N_n}\exp(-E_nN_n/T)}\nonumber\\
&&=  \frac{-\frac{d\left( \frac{1}{1+\exp(-E_n/T)} \right )}{
d(1/T)}}{\frac{1}{1+\exp(-E_n/T)}}
= -\frac{E_n}{\exp(E_n/T)+1} \hbox{(boson; neg. norm sq.)}\\
&&<E_nN_n>|_{fermion \, pos.} =
\frac{\sum_{N_n=0,1}E_nN_n\exp(-E_nN_n/T)}{\sum_{N_n=0,1}
\exp(-E_nN_n/T)}\nonumber\\
&=&  \frac{-\frac{d\left( {1+\exp(-E_n/T)} \right )}{
d(1/T)}}{{1+\exp(-E_n/T)}}
= \frac{E_n}{\exp(E_n/T)+1} \hbox{(fermion; pos. norm sq.)}\\
&&<E_nN_n>|_{fermion \, neg.} =
\frac{\sum_{N_n=0,1}(-1)^{N_n}E_nN_n\exp(-E_nN_n/T)}{\sum_{N_n=0,1}
(-1)^{N_n}\exp(-E_nN_n/T)}\nonumber\\
&&=  \frac{-\frac{d\left({1-\exp(-E_n/T)} \right )}{
d(1/T)}}{{1-\exp(-E_n/T)}}
= -\frac{E_n}{\exp(E_n/T)-1} \hbox{(fermion; neg. norm sq.)}
 \end{eqnarray}

We notice that - by accident - the contribution from 
a negative norm square fermion mode happens to be just 
the opposite of that of a positive norm square boson mode 
with the same energy $E_n$. And also the positive fermion 
mode contribution is just minus one time the negative 
boson contribution. Thus we can get the 
requirement for the theory of fermions and that of bosons
to provide the same average energy:
\begin{eqnarray}
\sum_{\substack{E_n's \hbox{for (pos.)fermions}\\ \hbox{ plus neg. bosons}}}\frac{E_n}
{\exp(E_n/T) +1} &=& \sum_{\substack{E_n's \hbox{for (pos.)bosons}\\ \hbox{ plus
neg. fermions}}}.   
\end{eqnarray} 

\subsection{Free Massless}

The simplest case to consider is the one in which 
both the fermions and the bosons - on their respective 
sides of the identification of the theories - are supposed
to be both free and massless relativistic particles. 
In this case - which is the one we shall keep to
in the present article - we introduce for definiteness 
an infra red cut off so that we get discretized momentum 
eigenstates, and the above $n$ now really becomes a 
pair of a discretized momentum $\vec{p}$ and an index denoting
the component, which means typically the vector or 
spinor index including also the family index, all put together
say to $t$, standing for the word ``total component'',
meaning that both family and genuine component is included. 
The number 
of possible values for this total component enumeration is 
of course for what we are indeed  obtaining restrictions for.
Let us therefore immediately define the four numbers
$$N_{t \;ferm\; pos.}= N_{families \; ferm \; pos.}*N_{c\; ferm \; pos},$$ 
$$N_{t \;ferm\; neg.}= N_{families\; ferm\; neg.}*N_{ c \; ferm \; neg.},$$
$$N_{t \;boson \; pos.}= N_{families \; boson \; pos.}*N_{c\; boson \; pos},$$
$$N_{t \;boson\; neg.}= N_{families \; boson \; neg.}*N_{c\; boson \; neg.},$$
to denote the total numbers of components of the respective 
types of particles w.r.t. statistics and normsquare sign.

One technique for calculating the integrals over the 
momentum space consists in first Taylor expanding the 
expressions to be integrated
\begin{eqnarray}
\frac{E_n}{\exp(E_n/T)-1}&=& \frac{E_n}{\exp(E_n/T)}*
(1+\exp(-E_n/T)+ \exp(-2E_n/T) + ...)\nonumber\\
&=& E_n\left ( \sum_{j=1,2,...}\exp(-jE_n/T) \right )\\
\frac{E_n}{\exp(E_n/T)+1}&=& \frac{E_n}{\exp(E_n/T)}*
(1-\exp(-E_n/T)+ \exp(-2E_n/T) - ...)\nonumber\\
&=& E_n\left ( \sum_{j=1,2,...}(-1)^{j-1}\exp(-jE_n/T) \right ),
\end{eqnarray}
and then using 
\begin{eqnarray}
&&\sum_{\vec{l} \in \hbox{integer lattice}}\exp(-j|\vec{l}*2\pi/ L|)=
\int \exp(-j|\vec{x} 2\pi/L|)d^{d_{spatial}}\vec{x}\\
&&= \left ( \frac{L}{2\pi *j}\right )^{d_{spatial}}\int 
\exp(-|\vec{x}|)d^{d_{spatial}}\vec{x}\\
&&= \left ( \frac{L}{2\pi *j}\right )^{d_{spatial}}
{\cal O}(d_{spatial} -1)\int_0^{\infty}\exp(-x)x^{d_{spatial}}dx\\
&&= \left ( \frac{L}{2\pi *j}\right )^{d_{spatial}}
{\cal O}(d_{spatial} -1)/d_{spatial}!.   
\end{eqnarray}    
Here we denoted the surface area of the unit sphere in 
$d_{spatial}$ dimensions by ${\cal O}(d_{spatial} -1)$ because 
this surface then has the dimension $d_{spatial} -1$.
In fact 
\begin{equation}
{\cal O}(d_{surface}) = \frac{2\pi^{d_{surface}/2}}
{\Gamma(d_{surface}/2)}.
\end{equation} 
We then finally shall use
\begin{eqnarray}
\zeta(d_{spatial})&=& \sum_{j=0,1,2,...}\frac{1}{j^{d_{spatial}}}.\\
\zeta(d_{spatial})\left (1- \frac{1}{2^{d_{spatial}}} \right )&=& 
\sum_{j=0,1,2,...}\frac{(-1)^j}{j^{d_{spatial}}}.
\end{eqnarray}

When we compare the different expressions for bosons 
versus for fermions, most factors drop out and 
the only important factor is the factor $
\left (1- \frac{1}{2^{d_{spatial}}} \right )$.
It is then easy to see that we obtain the 
extended Aratyn-Nielsen theorem:

\begin{eqnarray}
N_{t \; ferm\; pos.}+ N_{t\; boson \; neg.}&=& \frac{2^{d_{spatial}}}{
2^{d_{spatial}}-1}*(N_{t\; boson \; pos.} + N_{t\; ferm \; neg.}).\nonumber\\
\label{EANR} 
\end{eqnarray}

\subsection{Properties and Examples}

Let us first of all call attention to that this extended 
Aratyn-Nielsen theorem like the original one has the 
property of ``additivity'' meaning that if we have two 
cases of functioning bosonization - i.e. two cases 
of a system of fermions being equivalent to a system of bosons -
and thus by combining them formally a system with both 
sets of bosons making up its set of bosons and similarly 
construct a set of fermions by combing the fermions then 
the combined system will automatically - just algebraically -
come to obey the requirement from our theorem.

Let us also remark that the old Aratyn-Nielsen 
theorem\cite{Aratyn} 
just is the special case, in which there are no negative norm 
square components.

In the Bled workshop in 2015 \cite{Bled15} we presented 
speculations, that one could make a free massless case 
of bosonization/fermionization in an arbitrary number of 
dimensions. This attempt were indeed already strongly inspired 
from our theorem and counted just $2^{d_{spatial}}-1$ boson 
particle components and $2^{d_{spatial}}$ fermionic components.
There were no negative norm square components and the 
there suggested case of bosonization should thus be an 
example on the use of the ``old'' Aratyn-Nielsen theorem.
The ratio of the number $2^{d_{spatial}}$  of fermion 
components equivalent 
to $2^{d_{spatial}}-1$ bosonic components is namely of 
course just equal to $\frac{2^{d_{spatial}}}{2^{d_{spatial}}-1}$
as it should according to our theorem(s). The special feature 
of that proposal \cite{Bled15} was that we imagined having 
chosen such infrared cut off periodicity or antiperiodicity
conditions, that these (anti)periodicity conditions specified
the components of the fields. Indeed there were just one 
fermion component for each combination of a choice of 
periodicity versus antiperiodicity for each of the 
$d_{spatial}$ spatial dimensions. That makes up 
of course $ 2^{d_{spatial}}$ combinations of periodicity 
antiperiodicity choices and thus so many fermion 
components. Similarly almost all such combinations 
gave rise to a boson component, except that we deleted 
so to speak the boson components, that should have 
corresponded to being periodic in all $d_{spatial}$ 
coordinates (taken with infrared cut off). Thus there were
just $2^{d_{spatial}} -1$ boson components in the in this 
Bled proceeding speculated case of bosonization. 

\subsection{A speculative  semi-trivial example}
Starting from the example\cite{Bled15} we would now 
highly suggestively - but really a bit speculatively -
construct a not completely trivial although not so 
very physically interesting at first example with 
negative norm square components. Since we have anyway 
broken in this model full rotational invariance, it is no 
longer a catastrophe to treat one of coordinate axis -
say $x^1$ in 
a different way from the other ones.

We modify the model in the 2015 Bled proceedings by:
\begin{itemize}
\item On the fermionic side we take all the components 
specified by having odd momentum along say the $x^1$-axis
or equivalently have antiperiodic boundary condition 
in $x^1$ to have {\bf negative norm square}. They make up just
half - and thus $2^{d_{spatial}-1}$ - of all the 
fermionic components.
\item On the bosonic side we also change the norm-square to 
be negative for the components antiperiodic in the 
$x^1$-coordinate. This is for even more than half of the 
components in as far as it is again for $2^{d_{spatial} -1}$,
but now only out of the $2^{d_{spatial}}-1$ bosonic components. 
\end{itemize} 

Both of these two modifications have in the Fock-space 
the same effect in as far as they both just lead to 
shifting the norm square form positive to negative 
for all the states with the {\bf total $p^1$-momentum 
odd}. So the two modifications suggested for respectively
the bosons and the fermions seem to be the same one in the 
Fock space. At least speculatively then we expect, that 
the modified model will have functioning bosonization -
provided we trust that the original model from the Bled 2015
proceeding were indeed consistently a case of bosonization.

Now we want to test, if this suggestive speculative case 
of bosonization will obey our extended Aratyn-Nielsen 
requirement(\ref{EANR}):

We have in this modified model/case of bosonization:
\begin{itemize}
\item We are left with $2^{d_{spatial}-1}-1$ bosonic positive 
norm square components, i.e. 
 $N_{t \; boson \; pos.} =2^{d_{spatial}-1}-1$.
\item While $ 2^{d_{spatial}-1}$ of the bosonic components were
made to have negative norm squared. So $N_{t \; boson \; neg.}
= 2^{d_{spatial}-1}$.
\item Of the fermionic components $2^{d_{spatial} -1}$ remained
of positive norm-square; so $N_{t \; ferm \; pos.} = 2^{d_{spatial} -1}$.
\item Also $2^{d_{spatial}-1}$ components had the 
odd momentum in the $x^1$-direction and were made to have 
negative norm square. So $N_{t \; ferm \; neg.} = 2^{d_{spatial} -1 }$.\end{itemize}

Inserting these numbers of components into (\ref{EANR}) 
is easily seen to make it satisfied. The point really 
is, that we made the same number of boson components and 
of fermion components negative norm square. This sign of 
norm square in our formula makes them move from one side 
to the other, but since the two groups were of the same 
number at the end nothing were changed and the formula
still satisfied.

\section{Kovner...}
\label{Kovner}
{\bf \large Kovner and Kurzepa made 2+1}
The article by these authors \cite{KK} contains 
the expression 
\begin{eqnarray}
\psi_{\alpha}(x) &=& k\Lambda V_{\alpha}(x) \Phi(x) U_{\alpha}(x) 
\end{eqnarray}
for the fermion fields expressed in terms 
of the boson fields in their fermionization in 2+1
dimensions. Here the expressions $V_{\alpha}(x) $, 
$\Phi(x)$, and $U_{\alpha}(x)$ are exponentials of 
integrals over the boson field, which are 
indeed electromagnetic fields in 2+1 dimensions. 
The variants of expressions are denoted by the index
$\alpha$, which takes two values. There are 
thus (a priori) two complex fermion fields 
defined here.

\section{Match?}
\label{Match}
{\bf \large Does the Kovner Kurzepa Bosonization Match with the 
Aratyn-Nielsen Counting Rule?}

{\bf First look at number of hermitean counted fields:}
 Kovner and Kurzepa gets two complex meaning {\bf 4} real 
fermion fields $Re \psi_1(x)$, $Im \psi_1(x)$, $Re \psi_2(x)$,
and $Im \psi_2(x)$ out of the for the construction 
relevant boson-fields $A_1(x)$, $A_2(x)$, $\partial_iE_i
= \partial_1E_1 + \partial_2E_2$. This looks 
agreeing with the Aratyn Nielsen prediction that 
the ratio shall be 
\begin{equation}
\frac{\# bosons}{\# fermions} = \frac{2^{d_s}-1}{2^{d_s}}
=\frac{2^2 -1}{2^2} \hbox{for 
the spatial dimension being} d_s =d-1 = 2
\end{equation}
{\color{red} Four real fermion fields bosonize to 
three real boson-fields!} o.k.  
 

{\bf What about the conjugate momenta to the fields?}
While the fermion fields are normally each others 
conjugate variables(fields) in as far as they anticommute 
with each other having only  no-zero  anticommutators 
with themselves, the boson-fields typically are taken 
each to have associated an extra field - its 
conjugate - with which it does not commute, while 
of course any variable must commute with itself.
But a field, that depends on an x-point or on a momentum, 
need  NOT  to 
commute with itself, though.

But then the question: Shall we for bosons somehow 
also count the conjugate momentum fields, when we shall 
compare the number of fermion and boson fields 
equivalent through bosonization ?
For the fermions the conjugate fields are unavoidable 
already included into the set of fields describing the 
fermions, because the it is the field in question itself,
but for bosons we could easily get the number of fields 
{\em doubled}, if we include for each field also its 
conjugate.

{\bf Conjugate Momentum Fields NOT to be Included in Counting.}

Let us argue that it is enough in the counting to count the 
number 
of fields, from which you by Fourier resolution can extract 
the annihilation and creation operators needed to annihilate 
or create the particles, the species of which are  to be counted:
\begin{itemize}
\item Normally we could extract the conjugate field by 
differentiating w.r.t. to time the field because usually 
you can replace the fields and their conjugate by the 
fields and their time derivatives.
\item Using equations of motion these time derivatives can 
in turn be obtained by some way - also some sort of 
differentiation - from the field itself.
\item Thus at the end the information on the conjugate is 
extractable from the field itself!  
\end{itemize}

{\bf Further Support for NOT including also Conjugate Momentum 
Fields}

We could very easily construct linear (or more complicated)
combinations of boson fields and their conjugate fields.
Such combinations would like the fermion fields typically 
not commute/anticommute with themselves. 

So provided we can extra the particle creation and 
annihilation operators from the combined field we would 
have no rule to tell that we should include more. Thus 
we would need only the combined field, and with that 
rule have quite analogy to the fermion case.   

{\bf Meaning of NOT Counting also the Conjugate Field}

In $QED_3$ say $A_1(x)$ and $A_2(x)$ would be enough 
to represent both longitudinal and transversely polarized
photons. It would NOT be needed also to have the 
essentially conjugate electric fields $E_1(x)$ and 
$E_2(x)$. 

The field $\partial_iE_i$ is in fact the conjugate 
$A_0$ so that we - having the symmetry between a field 
and its conjugate, it being conjugate of its conjugate -
can consider that timelike photons are described by 
this $\partial_iE_i$ field combination. 

 \section{Particles}
\label{Particles}  

{\bf \large But in terms of Particles, How??}

Usually one thinks of electrodynamics in 2+1 
dimensions as having only one particle polarisation,
since there is only {\em one} transversely polarisation
for a 
photon. So seemingly only one component of boson. 
This transversely polarized photon is even 
its own antiparticle, so even the anti-particle 
is not new.

On the contrary the fermions after the fermionization
counts two complex fields meaning two different fermion
components ($\psi_1$ and $\psi_2$) each with an 
a priori different antiparticle in as far as the 
fields $\psi_1$ and $\psi_2$ both are complex(non-Hermitean).
That seems NOT to match!

{\color{red} Where have the two missing photon-polarizations
gone?} 

{\bf \large Suggestion for How 3 photons.}

To count independently both $A_i$ (i=1,2.) as 
real fields, we need to consider it that we have 
not only the transverse photon, but also 
a {\bf longitudinal photon} ! 

The third of the real fields $\partial_iE_i = div \vec{E}$
is actually the conjugate variable to the time component 
$A_0(x)$ of the fourcomponent photon field. So if we 
take it that conjugate or not does not matter it could 
correspond to the {\bf timelike polarized photon}.

This would mean that we could hope for interpreting the 
three photon polarizations as being 
\begin{itemize}
\item{1)} The transverse photon.
\item{2)} The longitudinal photon.
\item{3)} The time-like photon.
\end{itemize}

{\bf But the time like photon has wrong signature ?!}

{\bf \large Better Suggestion for the 3 particles ?}

To avoid the problem with the ltime-like photon 
form Lorentz invariance having the signature with 
negative norm square states we can instead take a 
further scalar. If so we could have 3 bosons corresponding
to the four (real) fermions. 

In any case if we want a fermion system with 
positive definite Hilbert space we better have the 
bosons also give positive definite Hilbert space if they
shall match in their Hilbert spaces. 
\section{Fields}
\label{Fields}
{\bf \large How to count Hermitean Boson fields ?}

To exercise we shall for the moment even 
begin with a 1+1 dimensional only right 
moving Hermitean field constructed as a superposition 
of momentum state creation $a^{\dagger}(p)$ and annihilation
operators $a(p)$ for say a series discretized momentum 
values, which we for ``elegance''( and later interest) shall 
take to 
be odd integers in some unit:
\begin{eqnarray}
\phi(x) &=& \sum_{p \hbox{ odd}, p>0} \sqrt{p} a(p)\exp{(ipx)} +
\sum_{p\hbox{ odd}, p<0}\sqrt{|p|}a^{\dagger}(|p|) \exp{(ipx)}\nonumber\\
&=& \sum_{p\hbox{ odd}} \sqrt{|p|} a(p),  
\end{eqnarray}
where we have put 
\begin{equation}
a(p) = a^{\dagger}(-p) \hbox{ for all the odd p} 
\end{equation} 

{\bf Properties of the Hermitean field}
A Hermitean field of the form (in 1+1 dimension say)
\begin{eqnarray}
\phi(x) &=& \sum_{p \hbox{ odd}, p>0} \sqrt{p} a(p)\exp{(ipx)} +
\sum_{p\hbox{ odd}, p<0}\sqrt{|p|}a^{\dagger}(|p|) \exp{(ipx)}\nonumber\\
&=& \sum_{p\hbox{ odd}} \sqrt{|p|} a(p)  
\end{eqnarray}
obeys
\begin{eqnarray}
\phi(x)^{\dagger} &=& \phi(x) \hbox{ (Hermiticity) }and 
\end{eqnarray}
\begin{eqnarray}
[ \phi (x), \phi (y)]& = & \sum_{p \hbox{ odd}} \sum_{p' \hbox{ odd}} 
\sqrt{|p|}\sqrt{|p'|}[a(p),a(p')]\exp{(ipx+ip'y)}\\
 &=&  \sum_{p \hbox{ odd}}p \exp{(ip(x-y))} 
 =2\pi \frac{d}{i d(x-y)}\delta (x-y)\\ 
& =& -i2\pi \partial \delta (x-y) 
\hbox{ (local commutation rule).}  
\end{eqnarray}

\section{New}
\label{New}
{\bf \large  New, Reduce the Kovner Kurzepa model.}

We claim, that in a way the Kovner and Kurzepa bosonization
in 2 + 1 dimensions has included a kind of ``funny extra 
bosonic degree of freedom'' the charge density compared 
to our own plan of doing a completely free model.

Really we want to say: In a truly free electrodynamics
``free $QED_3$'' (in 2 +1 dimensions) the divergence of 
the electric field is zero:
\begin{equation}
\partial_i E_i \approx 0 \hbox{  (on physical states).}
\end{equation}  
When we use $\approx$ instead of $=$ it is because 
we may need the divergence $\partial_i E_i$ as an operator
even though we may take it to be zero on the ``physical
states''.

{\bf Reduction of Kovner Kurzepa model w.r.t. degrees of freedom}

Inserting formally our claim of a constraint equation 
\begin{equation}
\partial_i E_i \approx 0 \hbox{  (on physical states).}
\end{equation} 
into the expressions of Kovner and Kurzepa
\begin{eqnarray}
   V_1(x)& =&-i\exp{(\frac{i}{2e}
\int (\theta (x-y) -\pi)\partial_iE_i)}\\
U_1(x) &=& \exp{(-\frac{i}{2e} \theta (y-x) \partial_iE_i) }
 \end{eqnarray}
we get 
\begin{eqnarray}
V_1(x)& \approx& -i\\
U_1(x) & \approx& 1. 
\end{eqnarray}
{\bf Using the constraint equation formally 
on Kovner and Kurzepa}
In Kovner and Kurzepa one finds
\begin{eqnarray}
\psi_{\alpha}(x) &=& k\Lambda V_{\alpha}(x) \Phi (x) U_{\alpha}(x)\\
\Phi (x) &=& \exp{(ie \int e_i(y-x) A_i(y) d^2y)}\hbox{;} e_i(y-x)
= \frac{y_i-x_i}{(y-x)^2}\\
 V_1(x)& =&-i\exp{(\frac{i}{2e}
\int (\theta (x-y) -\pi)\partial_iE_i)}\hbox{;}
 V_2(x) = -iV_1^{\dagger}(x)\\
U_1(x) &=& \exp{(-\frac{i}{2e} \theta (y-x) \partial_iE_i) }
\hbox{;} U_2(x) = V_1^{\dagger}(x) 
\end{eqnarray}
and thus with the constraint formally included
 \begin{eqnarray}
\psi_{2}(x)& \approx& i\psi_1(x) 
\end{eqnarray}
{\bf Our Constraint would Spoil Rotation Symmetry}
A constraint equation 
 \begin{eqnarray}
\psi_{2}(x)& \approx& i\psi_1(x) 
\end{eqnarray}
would {\em not} be consistent with the rotation symmetry 
and the transformation property for the fermion field
suggested in Kovner and Kurzepa
\begin{equation}
\psi_1 \rightarrow \exp{(i\phi/2)} \psi_1 ; \psi_2 \rightarrow 
\exp{(-i\phi/2)} \psi_2.
\end{equation}
So including the constraint would make the 
bosonization/fermionization become {\em non-rotational invariant}.
But it is our philosophy not to take that as a so serious 
problem, because it is in any case {\em impossible} to
get in a rotational invariant way spin 1/2 fermions 
from a purely bosonic theory with only integer spin! 

{\color{red} Rotation symmetry broken in reduced model!}

\section{Conclusion}
We have extended the previous ``Aratyn-Nielsen-thorem'' 
relating the number of degrees of freedom / number of 
components / number of particle (orthogonal) polarizations
for a set of bosons that by bosonization/fermionization 
is in correspondance with each other. The extension consists
in also allowing negative norm square single particle states.
We only considered yet the case of massless noninteracting 
both bosons and fermions, but expect that by thinking of the 
limit of small distances the relation of the theorem 
would also have to hold for massive particles. If there existed 
a common for both bosons and fermions weak interaction limit 
you would also expect that the noninteraction assumption could 
be avoided.  

The main result is the relation (\ref{EANR}): 
\begin{eqnarray}
N_{t \; ferm\; pos.}+ N_{t\; boson \; neg.}&=& \frac{2^{d_{spatial}}}{
2^{d_{spatial}}-1}*(N_{t\; boson \; pos.} + N_{t\; ferm \; neg.}),\nonumber
\end{eqnarray}
where the ``normal'' boson and fermion component numbers 
are denoted with $N_{t \; boson \; pos.}$ and $N_{t \; ferm \; pos.}$ 
respectively for bosons and for fermions, and where the 
corresponding numbers of components with negative norm 
square are   $N_{t \; boson \; neg.}$ and $N_{t \; ferm \; neg.}$.

We have also looked at some examples where one might apply 
and test our theorem, but the problem is that we do not 
know the higher dimensional examples so well. Basically 
the dimension limit where the examples basically stop is 
not high. Googling you find mainly at most 2+1. 
The case 3+1 is very rare.
\subsection{Outlook Dream}
Our motivation, which has not quite ran out to be realized 
yet is that we shall find in literature or  develop 
bosonization case(s)
for the dimensions of interest as dimension of the space time,
such as the experimental dimension 3+1 or the in the 
spin-charge-family theory practical starting dimension 
13 +1. That is to say we hope to find a set of 
boson fields that is equivalent to a set of fermion fields
in the bosonization way. If we have a valid theorem as 
the one we just extended we strictly speaking only need 
to know one side, i.e. either the bosons or the fermions,
because then we can calculate the number of components for 
the other side. Without the ``extension '' of our theorem 
it looks that the number of fermion components  must always 
be a number divisible by $2^{d_{spatial}}$, which e.g. for the 
case of the experimental dimension is $2^3 =8$. It makes 
it especially difficult to avoid the number of families 
being even, because if we think of Weyl fermions at least 
and even count real components so that we get twice as many 
as if we used complex components, we still need a multiplum 
of 2 families 
of Weyl particle. With Dirac fermions we could use up 
a factor 2 more and we would get no prediction than just the 
number of families being integer. But in the Standard model
we know that we have the weak interactions and the components 
put together to Dirac fermions have separate gauge quantum 
numbers are are hardly suitable for coming from the same 
fermionization. 

With an extended theorem relating the two sides fermions 
and bosons, however, the situation gets less clear and 
the hope for even getting somehow a phenomenologically good 
number is not excluded yet.

\section*{Acknowledgement}
One of us HBN thanks the Niels Bohr Institute for 
allowing him to stay as emeritus, and Matja\v z Breskvar, BS storitve d.o.o., for 
partly economic support to visit the Bled-conference.


\begin{thebibliography}{99}

\bibitem{axiom} Daniel Ranard, 
\textit{An introduction to rigorous formulations of quantum field theory},
Perimeter Institute for Theoretical Physics,
Advisors: Ryszard Kostecki and Lucien Hardy
May 2015.
%
\bibitem{axiom2} D. Iagolnitzer,
``Causality in Local Quantum Field Theory:
Some General Results'',
 Commun. Math. Phys. 144,235-255 (1992)
 Springer-Verlag 1992.
%
\bibitem{Aratyn} Aratyn, H.B. Nielsen, ``Constraints On 
Bosonization In Higher Dimensions'', NBI-HE-83-36, Conference:
C83-10-10.2 (Ahrenshoop Sympos.1983:0260), p.0260 Proceedings.
%
\bibitem{bosonization} Coleman, S. (1975). "Quantum 
sine-Gordon equation as the massive Thirring model" Physical 
Review D11 2088; Witten, E. (1984). "Non-abelian 
bosonization in two dimensions", Communications in 
Mathematical Physics 92 455-472.
E. Lieb, D. Mattis, J. Math. Phys. 6 (1965), 304; S. Coleman, Phys. Rev.
D11 (1975) 2088; A. Luther, I. Peschel, Phys. Rev. B12 (1975), 3908; S.
Mandelstam, Phys. Rev. D11 (1975), 3026.
%
%
%
\bibitem{IARD}  N.S. Manko\v c Bor\v stnik,  "Spin-charge-family theory is offering next step in
understanding elementary particles and fields and correspondingly universe", Proceedings to the 
Conference on Cosmology, Gravitational Waves and Particles, IARD conferences, Ljubljana, 6-9 
June 2016, The $10^{th}$ Biennial Conference on Classical and Quantum Relativistic Dynamics of
articles and Fields, J. Phys.: Conf. Ser. 845 012017
[arXiv:1607.01618v2], and the references therein.
%
\bibitem{Normas}
N. Manko\v c Bor\v stnik, "Spin connection as a 
              superpartner of a vielbein", Phys. Lett. B 292 (1992)  25-29.\\
 N. Manko\v c Bor\v stnik, "Spinor and vector representations in four dimensional Grassmann
              space", J. of Math. Phys. {\bf 34} (1993), 3731-3745.
 N.S. Manko\v c Bor\v stnik, "Matter-antimatter asymmetry in the 
{\it spin-charge-family} theory", Phys. Rev. {\bf D 91}  065004 (2015)
 [arxiv:1409.7791].\\ 
\bibitem{Normas2} 
 N.S. Manko\v c Bor\v stnik, "The explanation for the origin of the 
Higgs’s  scalar and for the Yukawa couplings by the {\it spin-charge-family} theory", 
 J.of Mod. Physics {\bf 6} (2015) 2244-2274, http://dx.org./10.4236/jmp.2015.615230
              [http://arxiv.org/abs/1409.4981], and the references therein.\\ 
%
 N.S. Manko\v c Bor\v stnik, D. Lukman, "Vector and scalar gauge fields with
              respect to $d=(3+1)$ in Kaluza-Klein theories and in the {\it spin-charge-family theory}",
              Eur. Phys. J. C {\bf 77} (2017) 231.\\
N.S. Manko\v c Bor\v stnik, H.B.F. Nielsen, "The spin-charge-family theory 
             offers understanding of the triangle anomalies cancellation in the standard model",
             Fortschrite der Physik, Progress of Physics (2017) 1700046.\\ 

\bibitem{KK}
  A.~Kovner and P.~S.~Kurzepa,
  ``Fermions from photons: Bosonization of QED in 
(2+1)-dimensions,''
  Int.\ J.\ Mod.\ Phys.\ A {\bf 9} (1994) 4669
  doi:10.1142/S0217751X94001862
  [hep-th/9306065].
\bibitem{KK2}
  A.~Kovner and P.~Kurzepa,
  ``Bosonization in (2+1)-dimensions 
without Chern-Simons attached,''
  Phys.\ Lett.\ B {\bf 321} (1994) 129
  doi:10.1016/0370-2693(94)90338-7
  [hep-th/9303145].

\bibitem{otherbosonization}
Andrea Cappelli, INFN Florence,
``Topological Insulators in 3D
and
Bosonization''
\bibitem{higher}Phys.Lett.B336:18-24,1994,
M. L{\"u}scher, Nuclear Physics B
Volume 326, Issue 3, 13 November 1989, Pages 557-582
Nuclear Physics B
``Bosonization in 2 + 1 dimensions''



\bibitem{Bled15}
 ``Fermionization in an Arbitrary Number of
Dimensions''
N.S. Manko\v c Bor\ stnik  and H.B.F. Nielsen 
in
Proceedings, 18th Workshop on ``What Comes Beyond the Standard 
Models?'' : Bled, Slovenia, July 11-19, 2015
Norma Susana Manko\v c Bor\v stnik (ed.) , Holger Bech Nielsen (ed.) 
, Dragan Lukman (ed.)(2015)

\bibitem{KR}  M. Kalb and Pierre Ramond, 
"Classical direct interstring action." 
Phys. Rev. D 9 (1974), 2273--2284.

   
\end{thebibliography}
\end{document}